\documentclass[twocolumn,10pt,aps,prl,superscriptaddress,showpacs,a4paper]{revtex4-1}

\usepackage[english]{babel}
\usepackage[utf8]{inputenc}

\usepackage{amsmath,amsfonts,amssymb}
\usepackage{graphicx}
\usepackage{hyperref}

\newcommand{\ie}{i.\,e.\ }
\newcommand{\cf}{cf.\ }

\newcommand{\dd}{\text{d}}

\newcommand{\re}{\operatorname{Re}}
\newcommand{\im}{\operatorname{Im}}

\newcommand{\ket}[1]{|{#1}\rangle}
\newcommand{\bra}[1]{\langle{#1}|}
\newcommand{\ketbra}[2]{|{#1}\rangle\langle{#2}|}
\newcommand{\ew}[1]{\langle{#1}\rangle}

\newcommand{\vct}[1]{\mathbf{#1}}
\newcommand{\vop}[1]{\hat{\vct{#1}}}

\newcommand{\abs}[1]{\lvert#1\rvert}
\newcommand{\tder}[2]{\tfrac{d^{#2}}{d {#1}^{#2}}}

\newcommand{\bkappa}{\boldsymbol{\kappa}}
\newcommand{\bepsilon}{\boldsymbol{\epsilon}}

\newcommand{\be}{\mathbf{e}}
\newcommand{\bk}{\mathbf{k}}

\newcommand{\bp}{\mathbf{p}}

\newcommand{\bd}{\mathbf{d}}
\newcommand{\bv}{\mathbf{v}}

\newcommand{\klam}{{\mathbf{k},\lambda}}
\newcommand{\klamp}{{\mathbf{k}',\lambda'}}

\begin{document}
\title{Will a decaying atom feel a friction force?}
\author{Matthias Sonnleitner}
\affiliation{School of Physics and Astronomy, University of Glasgow, Glasgow G12~8QQ, United Kingdom}
\author{Nils Trautmann}
\affiliation{School of Physics and Astronomy, University of Glasgow, Glasgow G12~8QQ, United Kingdom}
\affiliation{Institut f\"{u}r Angewandte Physik, Technische Universit\"{a}t Darmstadt, D-64289 Darmstadt, Germany}
\author{Stephen M. Barnett}
\affiliation{School of Physics and Astronomy, University of Glasgow, Glasgow G12~8QQ, United Kingdom}

\begin{abstract}
	We show how a simple calculation leads to the surprising result that an excited two-level atom moving through vacuum sees a tiny friction force of first order in~$v/c$. At first sight this seems to be in obvious contradiction to other calculations showing that the interaction with the vacuum does not change the velocity of an atom. It is yet more surprising that this change in the atom's momentum turns out to be a necessary result of energy and momentum conservation in special relativity.
\end{abstract}

%
\maketitle
%
%
Atoms moving through and interacting with electromagnetic fields experience velocity dependent forces. For laser fields this results in famous (sub-) Doppler cooling schemes which form the foundation of modern atom-optical experiments~\cite{stenholm1986semiclassical,cohen1998nobel,phillips1998nobel} while much earlier works by Einstein and Hopf already showed how such friction forces are necessary to discuss thermal radiation fields self-consistently~\cite{einstein1910statistische,einstein1917quantum,milonni1981quantum}. But these considerations also showed that no such forces will arise from the interaction with the vacuum. 

Here, however, we report how the simplest quantum-optical model, an initially excited atom moving through the vacuum, appears to be in contradiction with these results. Our starting point is the usual, non-relativistic atom-field Hamiltonian in the dipole approximation including the R\"ontgen term from which we first re-derive the decay rate from the excited state. Based on this well established framework we then go on to calculate the canonical and kinetic momenta of the atom and see that these are not constant in time, indicating the presence of a net force.

Such a force emanating from the vacuum is, of course, highly suspicious. Even more suspicious is the fact that this force has the form of a friction, that is, the change in momentum is proportional to the initial momentum. This suggests that there is no such force in the atom's rest frame and a co-moving observer will stay next to the particle while any other observer will see that the atom decelerates. Such a result would contradict the principle of relativity.

We find that only a proper consideration of energy and momentum conservation can solve this puzzle and that the momentum of the atom does change while its velocity remains constant.

For the sake of clarity and brevity we describe the atom as a two-level system, its dynamics will be calculated using time-dependent perturbation theory. This model is described by a Hamiltonian $H=H_0+H_\text{AF}$ where
\begin{equation}
	H_0=\frac{\vop{P}^2}{2M} + \hbar \omega_A \ketbra{e}{e}
	+ \sum_\klam \hbar \omega_k  \hat{a}_\klam^\dagger \hat{a}_\klam \,,
\end{equation}
describes the kinetic energy and dynamics of the atom with internal states $\ket{g}$ and $\ket{e}$ as well as the quantised electromagnetic field. The last term includes sums over all modes~$\bk$ and over the two polarisation states $\lambda=1,2$ while $\hat{a}_\klam$ is the usual annihilation operator for a photon in mode~$\bk$ and polarisation~$\lambda$. 

The interaction between the atom and the field is given in the dipole approximation where we include the R\"ontgen term~\cite{babiker1984theory,baxter1993canonical,lembessis1993theory,wilkens1994significance}
\begin{equation}\label{eq:_H_AF_original}
	H_\text{AF} = -\vop{d} \cdot \vct{E}^\perp - \tfrac{1}{2 M}\left[ \vop{P}\cdot (\vct{B} \times \vop{d}) + (\vct{B} \times \vop{d})\cdot \vop{P} \right] \,,
\end{equation}
with the electric dipole operator $\vop{d} = \vct{d} \left( \ketbra{e}{g} + \ketbra{g}{e}\right)$ where $\vct{d}=d \be_d$ and electric and magnetic fields given by
\begin{subequations}\label{eq:_EandB_vacuum}
\begin{align}
	\vct{E}^\perp(\vct{R}) &=	i \sum_{\klam} \mathcal{E}_k \bepsilon_\klam \left( \hat{a}_\klam e^{i \bk\cdot\vop{R}} - \hat{a}_\klam^\dagger e^{-i \bk\cdot\vop{R}} \right) \,,	
	\label{eq:_E_vacuum}
	\\
 	\vct{B}(\vct{R}) &=	\frac{i}{c} \sum_\klam \mathcal{E}_k \left(\bkappa \times \bepsilon_\klam\right) 
		\left( \hat{a}_\klam e^{i \bk\cdot\vop{R}} - \text{h.c.}\right) \,. \label{eq:_B_vacuum}
\end{align}
\end{subequations}
Here $\mathcal{E}_k=\sqrt{\hbar \omega_k/(2 \varepsilon_0  \mathcal{V})}$ is the electric field strength per photon in the quantisation volume $\mathcal{V}$ and $\bepsilon_\klam$ is a polarisation vector perpendicular to the mode propagating in a direction $\bkappa = \bk c/\omega_k$.

Using the fields from equ.~\eqref{eq:_EandB_vacuum} and $\exp(i \bk\cdot\vop{R}) \vop{P} =( \vop{P} - \hbar \bk) \exp(i \bk\cdot\vop{R}) $ we can rewrite the interaction Hamiltonian from equ.~\eqref{eq:_H_AF_original} in the rotating wave approximation,
\begin{equation}\label{eq:_H_AV}
	H_\text{AF} = i \hbar \sum_\klam \Omega_k \left(
			\hat{g}_\klam e^{i \bk\cdot\vop{R}} \ketbra{e}{g} \hat{a}_\klam
			- \text{h.c.} \right)
			 \,,
\end{equation}
with $\hbar \Omega_k := -d \mathcal{E}_k$ and a momentum-dependent coupling term
\begin{equation}\label{eq:_geom_term_g}
	\hat{g}_\klam := \bepsilon_\klam \cdot \be_d + 
	\tfrac{1}{Mc}\big(\vop{P}-\tfrac{\hbar\bk}{2}\big)\cdot \big( (\bkappa\times\bepsilon_\klam)\times\be_d\big) \,.
\end{equation}
The notation with a hat indicates that $\hat{g}_\klam$ depends on the momentum operator $\vop{P}$ and is thus an operator itself. More familiar results from a theory ignoring the R\"ontgen interaction term ${\sim\vct{B}\times\vct{d}}$ are recovered by setting $\hat{g}_\klam\rightarrow  \bepsilon_\klam \cdot \be_d$.

The R\"ontgen term included in equ.~\eqref{eq:_H_AF_original} describes the coupling between a moving electric dipole and a magnetic field~\cite{rontgen1888ueber}. Its presence leads to corrections ensuring the correct decay pattern of a moving atom~\cite{wilkens1994significance,cresser2003rate} and to a subtle difference between the canonical momentum $\vop{P}$ and the kinetic momentum $M \tder{t}{}\vop{R}$ where~\cite{babiker1984theory,baxter1993canonical,lembessis1993theory,barnett2010enigma,barnett2010resolution}
\begin{equation}\label{eq:_PkinPcan}
	M \tder{t}{}\vop{R} = \vop{P} - \vct{B}\times\vop{d} \,.
\end{equation}

A simple perturbative treatment suffices to demonstrate the problem. In this work we consider the evolution of an initially excited atom in an eigenstate of the canonical momentum operator in vacuum, \ie $\ket{\psi_e} := \ket{e, \bp_0, 0}$. The coupling~$H_\text{AF}$ initiates transitions to the ground state with reduced momentum and a photon of momentum~$\hbar \bk$, \ie $\ket{\psi_\klam}:=\ket{g,\bp_0-\hbar \bk, 1_\klam}$ with $\ket{1_\klam} := \hat{a}_\klam^\dagger \ket{0}$. These are, of course, eigenstates of $H_0$ with energies $\hbar \omega_e$ and $\hbar \omega_g(\bk)$, separated by $\widetilde{\omega}_\bk :=\omega_e - \omega_g(\bk)$,
\begin{equation}\label{eq:_omega_eg}
	 \widetilde{\omega}_\bk = \omega_A-\omega_k + \tfrac{1}{M}\bk\cdot\big(\bp_0-\hbar \bk/2\big) \,.
\end{equation}

The state at time~$t$ can thus be described by 
\begin{equation}\label{eq:_psit}
	\ket{\psi(t)} = c_e(t) e^{-i \omega_e t} \ket{\psi_e}
	+ \sum_{\klam} c_\klam(t) e^{-i \omega_g(\bk) t} \ket{\psi_\klam} \,,
\end{equation}
with $c_e(0)=1$ and $c_\klam(0)=0$. These amplitudes evolve as 
\begin{subequations} \label{eq:_cegdot}
\begin{align}
	\dot{c}_e(t) &= \sum_\klam \Omega_k g_\klam(\bp_0) c_\klam(t) e^{i \widetilde{\omega}_\bk t}		\, \label{eq:_cedot} \\
	\dot{c}_\klam(t) &= - \Omega_k g_\klam(\bp_0) c_e (t) e^{-i \widetilde{\omega}_\bk t} 	\, \label{eq:_cgdot}
\end{align}	
\end{subequations}
where $g_\klam(\bp)$ is the eigenvalue of $\hat{g}_\klam$ given in equ.~\eqref{eq:_geom_term_g}. We can integrate these to find that $\re c_\klam^\ast(t) \dot{c}_\klam(t)$ is
\begin{multline}\label{eq:_deltaapprox}
	\re \Omega_k^2 g_\klam^2(\bp_0) c_e(t) \int_0^t c_e^\ast(t') e^{-i \widetilde{\omega}_\bk(t-t')} \approx \\ \approx
	 \pi \Omega_k^2 g_\klam^2(\bp_0) \delta( \widetilde{\omega}_\bk) \,.
\end{multline}
Here we truncated $c_e(t) c_e^\ast(t') \approx 1 + \mathcal{O}(\Omega_k)$ and used $\re \int_0^t e^{-i x (t-t')} dt' \rightarrow \pi \delta(x)$ for $t\gg x$. Using these approximations we quickly obtain the decay rate $\Gamma = \tder{t}{} \sum_\klam \abs{c_\klam(t)}^2 = 2 \re \sum_\klam c_\klam^\ast(t) \dot{c}_\klam(t)$,
\begin{equation}\label{eq:_decay_rate}
	\Gamma = 2 \pi \sum_{\klam} \Omega_k^2 g_\klam^2(\bp_0) 
		\delta\left(\omega_A-\omega_k + \tfrac{\bk \cdot \bp_0 - \hbar k^2/2}{M} \right) \,
\end{equation}
which is what we would expect from Fermi's golden rule~\cite{wilkens1994significance,cresser2003rate}. Note that the delta function includes both the Doppler shift~$\bk\cdot\bp_0/M$ as well as the recoil shift~$\hbar k^2/(2M)$~\cite{barnett2010recoil}. A calculation of the total decay rate to first order $1/Mc^2$ is given in the appendix.

After calculating the decay rate it is a straightforward task to calculate the expectation value for the canonical momentum $\ew{\vct{P}(t)} = \bra{\psi(t)}\vop{P}\ket{\psi(t)}$. As $\vop{P}$ commutes with $H_0$ we see that $\ew{\vct{P}(t)} = \bp_0 \abs{c_e(t)}^2 + \sum_\klam (\bp_0-\hbar \bk) \abs{c_\klam(t)}^2$ such that $\tder{t}{}\ew{\vct{P}(t)} = - \sum_\klam \hbar \bk \tder{t}{} \abs{c_\klam(t)}^2$ or
\begin{equation}\label{eq:_Pcan_dot}
	\tder{t}{}\ew{\vct{P}(t)} = - 2 \pi \sum_\klam \hbar \bk \Omega_k^2 g_\klam^2(\bp_0) \delta(\widetilde{\omega}_\bk) \,.
\end{equation}
Comparing this to $\Gamma$ as given in equ.~\eqref{eq:_decay_rate} we see that the momentum change of the atom is simply given by the probability to emit a photon into a mode~$\bk$ multiplied by the momentum of such a photon, $\hbar \bk$. Let us stress again that the directional decay rate given in equ.~\eqref{eq:_decay_rate} is consistent with a first-order approximation of the emission pattern of a moving dipole~\cite{wilkens1994significance}. This result for the momentum change is thus required by momentum conservation.

As mentioned in equ.~\eqref{eq:_PkinPcan} kinetic and canonical momenta are different in the presence of the R\"ontgen term. This difference is given by a term
\begin{equation}\label{eq:_Bcrossd}
	\vct{B}\times\vop{d} = -\tfrac{i \hbar}{c} \sum_\klam \Omega_k \vct{b}_\klam \left( e^{i \bk\cdot\vop{R}} \ketbra{e}{g} \hat{a}_\klam
			- \text{h.c.}\right) \,,
\end{equation}
with $\vct{b}_\klam:= (\bkappa \times \bepsilon_\klam)\times \be_d$. In the notation of equ.~\eqref{eq:_psit} this is $\vct{B}\times\vop{d} = - \frac{i \hbar}{c} \sum \Omega_k \vct{b}_\klam \left( \ketbra{\psi_e}{\psi_\klam} - \ketbra{\psi_\klam}{\psi_e}\right)$. 

To see the difference in the change of these momenta we calculate the time derivative of
\begin{equation}
	\ew{\vct{B}\times\vct{d}} =
		-\tfrac{2 \hbar}{c} \im \sum_{\klam} \Omega_k \vct{b}_\klam 		
		c_e c_\klam^\ast e^{-i \widetilde{\omega}_\bk t} 
		\,.
\end{equation}
This derivative will consist of three terms (\cf equs.~\eqref{eq:_cegdot}): the first is proportional to $\Omega_k \dot{c}_e c_\klam^\ast \sim \Omega_k \Omega_{k'} c_\klamp c_\klam^\ast \sim \Omega_k^2 \Omega_{k'}^2$, it thus goes beyond our perturbation theory; the second is $c_e \dot{c}_\klam^\ast \exp(-i \omega_{eg} t) = - \Omega_k g_\klam(p_0-\hbar\bk/2) \abs{c_e}^2$, its imaginary part is thus zero; hence the only surviving term is
\begin{equation}
	\tder{t}{} \ew{\vct{B}\times\vct{d}} =
		\tfrac{2 \hbar}{c} \re \sum_{\klam} \Omega_k \vct{b}_\klam \widetilde{\omega}_\bk c_e c_\klam^\ast e^{-i \widetilde{\omega}_\bk t} \,.
\end{equation}
We recognise that this expression is quite similar to what we had in equ.~\eqref{eq:_deltaapprox}, but here we get $\tder{t}{} \ew{\vct{B}\times\vct{d}} \sim \sum \Omega_k^2 \vct{b}_\klam g_\klam(\bp_0) \widetilde{\omega}_\bk \delta( \widetilde{\omega}_\bk )$. This is an integral involving a delta function multiplied by its argument, $\tder{t}{} \ew{\vct{B}\times\vct{d}}$ thus \emph{vanishes} and the change of our kinetic momentum is equal to the change in canonical momentum given by equ.~\eqref{eq:_Pcan_dot}.

Calculating a value for $\tder{t}{}\ew{\vct{P}}$ from equ.~\eqref{eq:_Pcan_dot} involves rather straightforward, albeit lengthy, integration. In contrast to previous works~\cite{wilkens1994significance,cresser2003rate} we do not assume an infinitely heavy atom here, but keep recoil terms to first order of~$1/M$ as these appear in~$g_\klam^2(\bp_0)$ and the delta function. More details on this calculation are given in the appendix. In the end we find
\begin{equation}\label{eq:_Pdot_result}
	\tder{t}{}\ew{\vct{P}(t)} \simeq
		- \Gamma \frac{\hbar \omega_A}{M c^2} \bp_0 \,,
\end{equation}
where $\Gamma = \omega_A^3 d^2/(3 \pi \varepsilon_0 \hbar c^3)$ is the total decay rate from equ.~\eqref{eq:_decay_rate}. We see that the change in momentum is proportional to the atom's decay rate, the recoil energy and the initial momentum of the atom.

The momentum change given in equ.~\eqref{eq:_Pdot_result} is the main result and topic of this work. It shows that an atom spontaneously emitting a photon changes its average momentum at a rate proportional to its initial momentum~$\bp_0$. As mentioned in the introduction, such friction terms in the interaction between light and matter are not unusual. But here this friction comes from the interaction with vacuum and we see that there is no force in the atom's rest frame where $\bp_0=0$~\cite{[{This can also be derived using only the d.E interaction, c.~f. }]cohen1992atom}.

Equ.~\eqref{eq:_Pdot_result} seems to indicate that an observer co-moving with the initially excited atom will (on average) observe no change in the atom's motional state, as the photons are emitted in a random direction and the recoil adds up to zero. But any other observer will see that the atom changes its initial momentum $\bp_0\neq 0$. Thus we note that observers in two different reference frames record different physical behaviour, a clear contradiction to the principle of relativity. Note that this is different from the Unruh-Davis effect which applies to particles initially \emph{accelerating} through the vacuum~\cite{milonni1994quantum}.

Of course we know that the physics described by the Hamiltonian $H_0+H_\text{AF}$ will not include all relativistic effects, but yet we expect its results to be correct to first order in~$v/c$, especially after we included the R\"ontgen term which ensures that the emission pattern on the right hand side of equ.~\eqref{eq:_Pdot_result} is consistent with a moving dipole.

It is also of concern that the physics involved in this derivation was not at all complicated. Of course we made some assumptions, especially the dipole and rotating wave approximations and the truncation of the time-dependent perturbation theory, but these simplifications are standard textbook quantum optics~\cite{loudon2000quantum,sakurai2011modern} and should not lead to obviously wrong physical results. Another simplification was to put the atom into a momentum eigenstate, but our arguments are repeatable for more physical initial states which are superpositions of these.

The resolution to this puzzle comes when we recognise that, in general, a change in (kinetic) momentum $\bp = M \bv$ as given in equ.~\eqref{eq:_Pdot_result} is not only related to a change in velocity, but also to a change in mass as $\dot{\bp} = M \dot{\bv} + \dot{M} \bv$ (this relationship holds also in special relativity, at least to first order $v/c$). Further, we know that the total mass of an atom (here described as a dipole) is not just the sum of its constituents (electron and nucleus), but it is reduced by the internal binding energy $M = m_1 + m_2 - E_\text{binding}/c^2$. This mass defect is, of course, well known, but it is usually discussed in connection with particle and nuclear physics where the binding energies are considerably larger than in atomic physics.

In our case the decaying atom increases its binding energy by an amount~$\hbar \omega_A$, thus changing its mass at a rate $\Gamma$ as
\begin{equation}
	\dot{M} = - \Gamma \hbar \omega_A/c^2 \,.
\end{equation}
Such a change in mass is consistent with a change in momentum, $\dot{\bp}= - \Gamma \hbar \omega_A \bv_0/c^2 = -\Gamma \hbar \omega_A/(M c^2) \bp_0$, if the velocity remains unchanged, $\dot{\bv}=0$. The change in momentum of a decaying atom can therefore be explained by a change in its mass-energy while its \emph{velocity remains constant}. 

This also resolves the issue of two observers in different reference frames who now both see the same physics: an atom with constant velocity emitting a photon and thus losing energy as $\tder{t}{}\ew{H_\text{A}}=-\Gamma \hbar \omega_A$ with $H_\text{A}=\hbar \omega_A \ketbra{e}{e}$. This loss of energy is then also connected to a change in inertia and momentum. 

But this resolution brings up another question: how did the changing mass enter our description of an atom interacting with an electromagnetic field? It is certainly \emph{not included} in the Hamiltonian which can be derived by explicitly setting $M=m_1+m_2$~\cite{lembessis1993theory}. We can also directly calculate the change in the particle's velocity as $\tder{t}{2}\vop{R} = - [H, [H, \vop{R}]]/\hbar^2$ to find a nonzero acceleration because $\tder{t}{} \ew{\vct{B}\times\vct{d}} = 0$, \cf equ.~\eqref{eq:_PkinPcan}. A proper distinction between a change in momentum and an acceleration thus requires the derivation of a new atomic Hamiltonian which also includes a dynamical mass-energy term coupled to~$\vop{P}$. 

By virtue of the R\"ontgen term the Hamiltonian used here properly describes the physics involved in the emission of the photon and this is also what enters the calculation of $\tder{t}{} \ew{\vct{P}}$ in equ.~\eqref{eq:_Pcan_dot}. In the following short calculation we use a simple example to show that the proper description of the emission process is sufficient to derive the friction term.
\begin{figure}
	\centering
	\includegraphics[width=0.95\columnwidth]{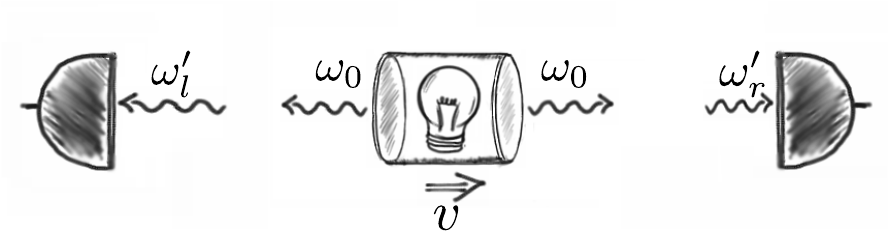}
	\caption{\label{fig:_towwaytorch}%
	A movable two-way emitter sending photons of equal energy to the left and to the right. An observer measuring the frequencies can use the Doppler shift to measure the relative velocity and will infer that the momentum of the device changed during the emission process.}
\end{figure}%

Figure~\ref{fig:_towwaytorch} shows a battery-powered device emitting a pair of photons with the same frequency in two opposite directions. If $E_i$ [$E_f$] denotes the total mass-energy of the apparatus before [after] the emission process we see that its internal energy changes as $\Delta E = E_f-E_i= -2 \hbar \omega_0$. The motion of the emitter remains constant by design as both photons carry equal momentum $\pm \hbar \omega_0/c$ in opposite directions. 

Let this apparatus now move at a speed~$v$ relative to two perfect detectors each measuring the frequency of one photon. Due to the Doppler shift these frequencies are different as $\omega_l'= \omega_0 \gamma (1-v/c)$ and $\omega_r'= \omega_0 \gamma (1+v/c)$ with $\gamma = (1-v^2/c^2)^{-1/2} \simeq 1 - v^2/(2 c^2)$. Using the available data the observer will construct the following relations of energy-momentum conservation
\begin{subequations}\label{eq:_TwoTorchExample_enmom2}
	\begin{align}
		E_i' &= E_f' + \hbar ( \omega_l' + \omega_r')		\\
		p_i' &= p_f' + \hbar ( - \omega_l' + \omega_r')/c
	\end{align}
\end{subequations}
such that $\Delta E'=E_f'-E_i'= - 2 \hbar \omega_0 \gamma$ and $\Delta p'=p_f'-p_i'=- 2 \hbar \omega_0 \gamma v/c^2$. We thus see that the observer would calculate a change in the momentum of the apparatus even though there is no change in its velocity~\cite{[{Einstein himself considered a similar setup in more detail when he discussed whether a change in internal energy is related to a change in inertia. }]einstein1905tragheit}.

This simple example illustrates how the Doppler effect directly links the change in internal energy during an emission process to a friction-like change in the momentum. Note again that although this momentum change is real it is not connected to an acceleration, the motion of the object remains unchanged. In our case of the decaying atom the presence of the R\"ontgen term in equ.~\eqref{eq:_H_AF_original} led to a correct emission pattern in equ.~\eqref{eq:_decay_rate} which in turn was all we needed to calculate the momentum change of the atom in equ.~\eqref{eq:_Pcan_dot}. Our calculations were thus similar to the energy-momentum balance in equs.~\eqref{eq:_TwoTorchExample_enmom2} and also gave the same result $\Delta p' = \Delta E' v/c^2$.

While preparing this letter we were made aware of a work by Wei Guo discussing the same setup, but omitting the R\"{o}ntgen interaction~\cite{guo2012radiative}. Unsurprisingly this gives a different form to the friction term derived here and, importantly, one that which would truly violate the principle of relativity, as observed by the author.

In conclusion we have shown how a simple calculation of an excited two-level atom in vacuum results in a momentum change proportional to the initial velocity of the atom, \cf equ.~\eqref{eq:_Pdot_result}. If this momentum change were linked to a true deceleration of the atom, it would be in contradiction to the principle of relativity. This riddle is solved, however, by recognising the change in the internal energy (mass) of the atom as the source of the momentum change. Thus, the momentum of the atom changes, but its velocity remains constant.

Let us stress that this effect is not, intrinsically, a quantum optical phenomenon and is also not limited to the special case of a decaying atom in vacuum. Rather it is a general phenomenon occurring whenever a particle changes its internal energy, for instance, due to stimulated absorption and emission or, more subtle, due to Stark or Zeeman shifts. In these cases with external fields and resulting (friction) forces this effect is concealed within the stronger interactions leading to a true acceleration.

In contrast to nuclear or particle physics, the energies involved in atom optics are far below the atom's rest mass. It is therefore not surprising that this subtle difference between momentum change and acceleration has not been measured yet. Ion traps have been used to measure nuclear masses to a precision of~$1-10 \text{ keV}$~\cite{blaum2013precision} which is roughly three orders of magnitude more than the mass defect due to atomic binding energies, $M_e-M_g=\hbar \omega_A/c^2 \approx 1\text{ eV}$.

Finally we would like to point out again that the current Hamiltonian describing a dipole in an external field does not include a dynamic mass. Although the calculation of the momentum change~$\ew{\dot{\vct{P}}}$ bypasses this insufficiency, the result obtained for the actual acceleration~$\ew{\ddot{\vct{R}}}$ is flawed. This calls for a small correction of the atomic Hamiltonian to include the coupling between the momentum and a change in internal energy which will be the topic of a future work.

Answering the question in the title: we have shown that, yes, a decaying atom sees a force resembling friction. But this force is a change in momentum due to a change in internal mass-energy and is not connected to decelerated motion.
%
%
\begin{acknowledgments}
We thank M.~Babiker, J.~Cresser and H.~Ritsch for stimulating discussions and gratefully acknowledge funding by the Austrian Science Fund FWF (J~3703-N27), the DFG as part of the CRC~1119 CROSSING, the EPSRC through QuantIC EP/M01326X/1 and the Royal Society.
\end{acknowledgments}
%
%

\begin{thebibliography}{22}%
\makeatletter
\providecommand \@ifxundefined [1]{%
 \@ifx{#1\undefined}
}%
\providecommand \@ifnum [1]{%
 \ifnum #1\expandafter \@firstoftwo
 \else \expandafter \@secondoftwo
 \fi
}%
\providecommand \@ifx [1]{%
 \ifx #1\expandafter \@firstoftwo
 \else \expandafter \@secondoftwo
 \fi
}%
\providecommand \natexlab [1]{#1}%
\providecommand \enquote  [1]{``#1''}%
\providecommand \bibnamefont  [1]{#1}%
\providecommand \bibfnamefont [1]{#1}%
\providecommand \citenamefont [1]{#1}%
\providecommand \href@noop [0]{\@secondoftwo}%
\providecommand \href [0]{\begingroup \@sanitize@url \@href}%
\providecommand \@href[1]{\@@startlink{#1}\@@href}%
\providecommand \@@href[1]{\endgroup#1\@@endlink}%
\providecommand \@sanitize@url [0]{\catcode `\\12\catcode `\$12\catcode
  `\&12\catcode `\#12\catcode `\^12\catcode `\_12\catcode `\%12\relax}%
\providecommand \@@startlink[1]{}%
\providecommand \@@endlink[0]{}%
\providecommand \url  [0]{\begingroup\@sanitize@url \@url }%
\providecommand \@url [1]{\endgroup\@href {#1}{\urlprefix }}%
\providecommand \urlprefix  [0]{URL }%
\providecommand \Eprint [0]{\href }%
\providecommand \doibase [0]{http://dx.doi.org/}%
\providecommand \selectlanguage [0]{\@gobble}%
\providecommand \bibinfo  [0]{\@secondoftwo}%
\providecommand \bibfield  [0]{\@secondoftwo}%
\providecommand \translation [1]{[#1]}%
\providecommand \BibitemOpen [0]{}%
\providecommand \bibitemStop [0]{}%
\providecommand \bibitemNoStop [0]{.\EOS\space}%
\providecommand \EOS [0]{\spacefactor3000\relax}%
\providecommand \BibitemShut  [1]{\csname bibitem#1\endcsname}%
\let\auto@bib@innerbib\@empty
\bibitem [{\citenamefont {Stenholm}(1986)}]{stenholm1986semiclassical}%
  \BibitemOpen
  \bibfield  {author} {\bibinfo {author} {\bibfnamefont {S.}~\bibnamefont
  {Stenholm}},\ }\href@noop {} {\bibfield  {journal} {\bibinfo  {journal} {Rev.
  Mod. Phys.}\ }\textbf {\bibinfo {volume} {58}},\ \bibinfo {pages} {699}
  (\bibinfo {year} {1986})}\BibitemShut {NoStop}%
\bibitem [{\citenamefont {Cohen-Tannoudji}(1998)}]{cohen1998nobel}%
  \BibitemOpen
  \bibfield  {author} {\bibinfo {author} {\bibfnamefont {C.~N.}\ \bibnamefont
  {Cohen-Tannoudji}},\ }\href@noop {} {\bibfield  {journal} {\bibinfo
  {journal} {Rev. Mod. Phys.}\ }\textbf {\bibinfo {volume} {70}},\ \bibinfo
  {pages} {707} (\bibinfo {year} {1998})}\BibitemShut {NoStop}%
\bibitem [{\citenamefont {Phillips}(1998)}]{phillips1998nobel}%
  \BibitemOpen
  \bibfield  {author} {\bibinfo {author} {\bibfnamefont {W.~D.}\ \bibnamefont
  {Phillips}},\ }\href@noop {} {\bibfield  {journal} {\bibinfo  {journal} {Rev.
  Mod. Phys.}\ }\textbf {\bibinfo {volume} {70}},\ \bibinfo {pages} {721}
  (\bibinfo {year} {1998})}\BibitemShut {NoStop}%
\bibitem [{\citenamefont {Einstein}\ and\ \citenamefont
  {Hopf}(1910)}]{einstein1910statistische}%
  \BibitemOpen
  \bibfield  {author} {\bibinfo {author} {\bibfnamefont {A.}~\bibnamefont
  {Einstein}}\ and\ \bibinfo {author} {\bibfnamefont {L.}~\bibnamefont
  {Hopf}},\ }\href@noop {} {\bibfield  {journal} {\bibinfo  {journal} {Ann.
  Phys. (Berlin)}\ }\textbf {\bibinfo {volume} {338}},\ \bibinfo {pages} {1105}
  (\bibinfo {year} {1910})}\BibitemShut {NoStop}%
\bibitem [{\citenamefont {Einstein}(1917)}]{einstein1917quantum}%
  \BibitemOpen
  \bibfield  {author} {\bibinfo {author} {\bibfnamefont {A.}~\bibnamefont
  {Einstein}},\ }\href@noop {} {\bibfield  {journal} {\bibinfo  {journal}
  {Phys. Z}\ }\textbf {\bibinfo {volume} {18}},\ \bibinfo {pages} {121}
  (\bibinfo {year} {1917})}\BibitemShut {NoStop}%
\bibitem [{\citenamefont {Milonni}(1981)}]{milonni1981quantum}%
  \BibitemOpen
  \bibfield  {author} {\bibinfo {author} {\bibfnamefont {P.~W.}\ \bibnamefont
  {Milonni}},\ }\href@noop {} {\bibfield  {journal} {\bibinfo  {journal} {Am.
  J. Phys.}\ }\textbf {\bibinfo {volume} {49}},\ \bibinfo {pages} {177}
  (\bibinfo {year} {1981})}\BibitemShut {NoStop}%
\bibitem [{\citenamefont {Babiker}(1984)}]{babiker1984theory}%
  \BibitemOpen
  \bibfield  {author} {\bibinfo {author} {\bibfnamefont {M.}~\bibnamefont
  {Babiker}},\ }\href@noop {} {\bibfield  {journal} {\bibinfo  {journal} {J.
  Phys. B}\ }\textbf {\bibinfo {volume} {17}},\ \bibinfo {pages} {4877}
  (\bibinfo {year} {1984})}\BibitemShut {NoStop}%
\bibitem [{\citenamefont {Baxter}\ \emph {et~al.}(1993)\citenamefont {Baxter},
  \citenamefont {Babiker},\ and\ \citenamefont {Loudon}}]{baxter1993canonical}%
  \BibitemOpen
  \bibfield  {author} {\bibinfo {author} {\bibfnamefont {C.}~\bibnamefont
  {Baxter}}, \bibinfo {author} {\bibfnamefont {M.}~\bibnamefont {Babiker}}, \
  and\ \bibinfo {author} {\bibfnamefont {R.}~\bibnamefont {Loudon}},\
  }\href@noop {} {\bibfield  {journal} {\bibinfo  {journal} {Phys. Rev. A}\
  }\textbf {\bibinfo {volume} {47}},\ \bibinfo {pages} {1278} (\bibinfo {year}
  {1993})}\BibitemShut {NoStop}%
\bibitem [{\citenamefont {Lembessis}\ \emph {et~al.}(1993)\citenamefont
  {Lembessis}, \citenamefont {Babiker}, \citenamefont {Baxter},\ and\
  \citenamefont {Loudon}}]{lembessis1993theory}%
  \BibitemOpen
  \bibfield  {author} {\bibinfo {author} {\bibfnamefont {V.~E.}\ \bibnamefont
  {Lembessis}}, \bibinfo {author} {\bibfnamefont {M.}~\bibnamefont {Babiker}},
  \bibinfo {author} {\bibfnamefont {C.}~\bibnamefont {Baxter}}, \ and\ \bibinfo
  {author} {\bibfnamefont {R.}~\bibnamefont {Loudon}},\ }\href@noop {}
  {\bibfield  {journal} {\bibinfo  {journal} {Phys. Rev. A}\ }\textbf {\bibinfo
  {volume} {48}},\ \bibinfo {pages} {1594} (\bibinfo {year}
  {1993})}\BibitemShut {NoStop}%
\bibitem [{\citenamefont {Wilkens}(1994)}]{wilkens1994significance}%
  \BibitemOpen
  \bibfield  {author} {\bibinfo {author} {\bibfnamefont {M.}~\bibnamefont
  {Wilkens}},\ }\href@noop {} {\bibfield  {journal} {\bibinfo  {journal} {Phys.
  Rev. A}\ }\textbf {\bibinfo {volume} {49}},\ \bibinfo {pages} {570} (\bibinfo
  {year} {1994})}\BibitemShut {NoStop}%
\bibitem [{\citenamefont {R{\"o}ntgen}(1888)}]{rontgen1888ueber}%
  \BibitemOpen
  \bibfield  {author} {\bibinfo {author} {\bibfnamefont {W.~C.}\ \bibnamefont
  {R{\"o}ntgen}},\ }\href@noop {} {\bibfield  {journal} {\bibinfo  {journal}
  {Ann. Phys. (Berlin)}\ }\textbf {\bibinfo {volume} {271}},\ \bibinfo {pages}
  {264} (\bibinfo {year} {1888})}\BibitemShut {NoStop}%
\bibitem [{\citenamefont {Cresser}\ and\ \citenamefont
  {Barnett}(2003)}]{cresser2003rate}%
  \BibitemOpen
  \bibfield  {author} {\bibinfo {author} {\bibfnamefont {J.~D.}\ \bibnamefont
  {Cresser}}\ and\ \bibinfo {author} {\bibfnamefont {S.~M.}\ \bibnamefont
  {Barnett}},\ }\href@noop {} {\bibfield  {journal} {\bibinfo  {journal} {J.
  Phys. B}\ }\textbf {\bibinfo {volume} {36}},\ \bibinfo {pages} {1755}
  (\bibinfo {year} {2003})}\BibitemShut {NoStop}%
\bibitem [{\citenamefont {Barnett}\ and\ \citenamefont
  {Loudon}(2010)}]{barnett2010enigma}%
  \BibitemOpen
  \bibfield  {author} {\bibinfo {author} {\bibfnamefont {S.~M.}\ \bibnamefont
  {Barnett}}\ and\ \bibinfo {author} {\bibfnamefont {R.}~\bibnamefont
  {Loudon}},\ }\href@noop {} {\bibfield  {journal} {\bibinfo  {journal} {Proc.
  R. Soc. London, Ser. A}\ }\textbf {\bibinfo {volume} {368}},\ \bibinfo
  {pages} {927} (\bibinfo {year} {2010})}\BibitemShut {NoStop}%
\bibitem [{\citenamefont
  {Barnett}(2010{\natexlab{a}})}]{barnett2010resolution}%
  \BibitemOpen
  \bibfield  {author} {\bibinfo {author} {\bibfnamefont {S.~M.}\ \bibnamefont
  {Barnett}},\ }\href@noop {} {\bibfield  {journal} {\bibinfo  {journal} {Phys.
  Rev. Lett.}\ }\textbf {\bibinfo {volume} {104}},\ \bibinfo {pages} {070401}
  (\bibinfo {year} {2010}{\natexlab{a}})}\BibitemShut {NoStop}%
\bibitem [{\citenamefont {Barnett}(2010{\natexlab{b}})}]{barnett2010recoil}%
  \BibitemOpen
  \bibfield  {author} {\bibinfo {author} {\bibfnamefont {S.~M.}\ \bibnamefont
  {Barnett}},\ }\href@noop {} {\bibfield  {journal} {\bibinfo  {journal} {J.
  Mod. Opt.}\ }\textbf {\bibinfo {volume} {57}},\ \bibinfo {pages} {1445}
  (\bibinfo {year} {2010}{\natexlab{b}})}\BibitemShut {NoStop}%
\bibitem [{\citenamefont {Cohen-Tannoudji}\ \emph {et~al.}(1992)\citenamefont
  {Cohen-Tannoudji}, \citenamefont {Dupont-Roc}, \citenamefont {Grynberg},\
  and\ \citenamefont {Thickstun}}]{cohen1992atom}%
  \BibitemOpen
  \bibfield  {author} {\bibinfo {author} {\bibfnamefont {C.}~\bibnamefont
  {Cohen-Tannoudji}}, \bibinfo {author} {\bibfnamefont {J.}~\bibnamefont
  {Dupont-Roc}}, \bibinfo {author} {\bibfnamefont {G.}~\bibnamefont
  {Grynberg}}, \ and\ \bibinfo {author} {\bibfnamefont {P.}~\bibnamefont
  {Thickstun}},\ }\href@noop {} {\emph {\bibinfo {title} {Atom-photon
  interactions: basic processes and applications}}}\ (\bibinfo  {publisher}
  {Wiley \& Sons},\ \bibinfo {year} {1992})\BibitemShut {NoStop}%
\bibitem [{\citenamefont {Milonni}(1994)}]{milonni1994quantum}%
  \BibitemOpen
  \bibfield  {author} {\bibinfo {author} {\bibfnamefont {P.~W.}\ \bibnamefont
  {Milonni}},\ }\href@noop {} {\emph {\bibinfo {title} {The Quantum Vacuum}}}\
  (\bibinfo  {publisher} {Academic Press},\ \bibinfo {year} {1994})\BibitemShut
  {NoStop}%
\bibitem [{\citenamefont {Loudon}(2000)}]{loudon2000quantum}%
  \BibitemOpen
  \bibfield  {author} {\bibinfo {author} {\bibfnamefont {R.}~\bibnamefont
  {Loudon}},\ }\href@noop {} {\emph {\bibinfo {title} {The quantum theory of
  light}}},\ \bibinfo {edition} {3rd}\ ed.\ (\bibinfo  {publisher} {Oxford
  University Press},\ \bibinfo {year} {2000})\BibitemShut {NoStop}%
\bibitem [{\citenamefont {Sakurai}\ and\ \citenamefont
  {Napolitano}(2011)}]{sakurai2011modern}%
  \BibitemOpen
  \bibfield  {author} {\bibinfo {author} {\bibfnamefont {J.~J.}\ \bibnamefont
  {Sakurai}}\ and\ \bibinfo {author} {\bibfnamefont {J.}~\bibnamefont
  {Napolitano}},\ }\href@noop {} {\emph {\bibinfo {title} {Modern quantum
  mechanics}}},\ \bibinfo {edition} {2nd}\ ed.\ (\bibinfo {year}
  {2011})\BibitemShut {NoStop}%
\bibitem [{\citenamefont {Einstein}(1905)}]{einstein1905tragheit}%
  \BibitemOpen
  \bibfield  {author} {\bibinfo {author} {\bibfnamefont {A.}~\bibnamefont
  {Einstein}},\ }\href@noop {} {\bibfield  {journal} {\bibinfo  {journal} {Ann.
  Phys. (Berlin)}\ }\textbf {\bibinfo {volume} {323}},\ \bibinfo {pages} {639}
  (\bibinfo {year} {1905})}\BibitemShut {NoStop}%
\bibitem [{\citenamefont {Guo}(2012)}]{guo2012radiative}%
  \BibitemOpen
  \bibfield  {author} {\bibinfo {author} {\bibfnamefont {W.}~\bibnamefont
  {Guo}},\ }\href@noop {} {\bibfield  {journal} {\bibinfo  {journal}
  {arXiv:1204.6646}\ } (\bibinfo {year} {2012})}\BibitemShut {NoStop}%
\bibitem [{\citenamefont {Blaum}\ \emph {et~al.}(2013)\citenamefont {Blaum},
  \citenamefont {Dilling},\ and\ \citenamefont
  {N{\"o}rtersh{\"a}user}}]{blaum2013precision}%
  \BibitemOpen
  \bibfield  {author} {\bibinfo {author} {\bibfnamefont {K.}~\bibnamefont
  {Blaum}}, \bibinfo {author} {\bibfnamefont {J.}~\bibnamefont {Dilling}}, \
  and\ \bibinfo {author} {\bibfnamefont {W.}~\bibnamefont
  {N{\"o}rtersh{\"a}user}},\ }\href@noop {} {\bibfield  {journal} {\bibinfo
  {journal} {Phys. Scr.}\ }\textbf {\bibinfo {volume} {2013}},\ \bibinfo
  {pages} {014017} (\bibinfo {year} {2013})}\BibitemShut {NoStop}%
\end{thebibliography}
%
%
%
\newpage
\onecolumngrid
\appendix
\numberwithin{equation}{section}
\section[Supplementary material]{Appendix: Calculation of decay rate and momentum change integrals}
\setcounter{equation}{0}
\setcounter{section}{1}
In equs.~(11) and~(12) we show that the decay rate and the change in momentum of an excited atom with initial momentum~$\bp_0$ are given by
\begin{gather}
	\Gamma = 2 \pi \sum_{\klam} \Omega_k^2 g_\klam^2(\bp_0) 
		\delta\big(\omega_A-\omega_k + \tfrac{1}{M}\bk\cdot(\bp_0-\hbar \bk/2)\big) \,,
		\label{eq:_decay_rate_appendix} \\
	\tder{t}{}\ew{\vct{P}(t)} = - 2 \pi \sum_\klam \hbar \bk \Omega_k^2 g_\klam^2(\bp_0) 
		\delta\big(\omega_A-\omega_k + \tfrac{1}{M}\bk\cdot(\bp_0-\hbar \bk/2)\big) \,.
		\label{eq:_Pcan_dot_appendix}
\end{gather}
We shall assume that the atom is heavy and expand only to first order in $(Mc)^{-1}$ such that, for $\bd = d \be_d$,
\begin{equation}
	d^2 g_\klam^2(\vct{p_0}) \simeq 
		(\bd\cdot\bepsilon_\klam)^2 +\tfrac{2}{Mc} (\bd \cdot\bepsilon_\klam) \big(\bp_0-\tfrac{\hbar \omega}{2 c}\bkappa\big)\cdot\big( (\bkappa\times\bepsilon_\klam)\times\bd\big) \,.
\end{equation}
The two polarisation directions~$\bepsilon_\klam$ and the unit wave vector~$\bkappa = \bk c/\omega_k$ are mutually orthonormal. Hence summing over the polarisations $\lambda=1,2$ we get $\sum_\lambda (\bd\cdot\bepsilon_\klam)^2 = d^2 - (\bd \cdot\bkappa)^2$ and
\begin{equation}
	\sum_\lambda (\bd\cdot\bepsilon_\klam) \vct{a}\cdot\big( (\bkappa\times\bepsilon_\klam)\times\bd\big)
		=
		(\bd \cdot \bkappa)(\vct{a}\cdot\bd) - d^2 (\vct{a}\cdot\bkappa)
\end{equation}
for any vector $\vct{a}=(\vct{a}\cdot\bkappa) \bkappa + \sum_\lambda (\vct{a}\cdot\bepsilon_\klam) \bepsilon_\klam$, but here specifically for $\vct{a}=\tfrac{2}{M c} \big(\bp_0-\tfrac{\hbar \omega}{2 c}\bkappa\big)$.

If we now change the sum over~$\bk$ to an integral of continuous modes $\sum_\bk \Omega_k^2 \rightarrow \tfrac{d^2}{2 (2\pi c)^3 \hbar \varepsilon_0} \int \dd\bkappa \int \dd\omega\; \omega^3$ we obtain
\begin{gather}
	\Gamma = \frac{\pi}{(2 \pi c)^3 \hbar \varepsilon_0} \int_{4\pi} \dd \bkappa \int_0^\infty \dd \omega\; f_\Gamma(\omega) 
		\delta\big(\omega_A-\omega_k + \tfrac{1}{M}\bk\cdot(\bp_0-\hbar \bk/2)\big) \\
	\tder{t}{}\ew{\vct{P}(t)} = - \frac{\pi}{(2 \pi c)^3 c \varepsilon_0} \int_{4\pi} \dd \bkappa \int_0^\infty \dd \omega\; \bkappa \; f_{\dot{\vct{P}}}(\omega) \delta\big(\omega_A-\omega_k + \tfrac{1}{M}\bk\cdot(\bp_0-\hbar \bk/2)\big) 
\end{gather}
with the solid angle integral $\int_{4\pi} \dd\bkappa := \int_{-1}^1 \dd\cos\theta \int_0^{2\pi} \dd\phi$ for $\bkappa=(\sin\theta \cos \phi, \sin\theta \sin\phi, \cos\theta)$ and 
\begin{equation}
	f_\Gamma(\omega) \simeq \omega^3 \left(1+\tfrac{\hbar \omega}{Mc^2}\right) \left(d^2 - (\bd \cdot \bkappa)^2\right) 
					+ \tfrac{2}{M c} \omega^3 \left( (\bp_0\cdot\bd) (\bd\cdot \bkappa) - \bp_0\cdot\bkappa d^2 \right)
\end{equation}
and $f_{\dot{\vct{P}}}(\omega) = \omega f_\Gamma(\omega)$.

Generally we have $\int_a^b f(x) \delta(h(x)) \dd x = \sum_{x_0} f(x_0)/\abs{h'(x_0)}$ for smooth functions $f$ and $h$ where $x_0$ are all zeros of $h$ within the interval $(a,b)$ and $h'(x_0)\neq 0$. In our case $h(\omega) \equiv \widetilde{\omega}_\bk = \omega_A-\omega (1- \bkappa\cdot\bp_0/(Mc)) - \hbar \omega^2/(2 M c^2)$ has only one positive root at
\begin{equation}
	\omega_+ = \frac{M c^2}{\hbar}\left[ -\left(1-\frac{\bkappa\cdot\bp_0}{Mc}\right)
					+ \sqrt{\left(1-\frac{\bkappa\cdot\bp_0}{Mc}\right)^2 + 2\frac{\hbar \omega_A}{M c^2}}\right]					
			= \omega_A + \frac{\omega_A}{Mc}\left(\bkappa\cdot\bp_0 - \frac{\hbar \omega_A}{2c}\right) 
			+ \mathcal{O}\Big( \big(\tfrac{\hbar \omega_A}{Mc^2}\big)^2 \Big)\,,
\end{equation}
with $\abs{h'(\omega_+)}\simeq 1-\bkappa\cdot\bp_0/(Mc) + \hbar \omega_A/(Mc^2)$. We thus expand, again to first order in $(Mc)^{-1}$,
\begin{equation}
	\frac{f(\omega_+)}{\abs{h'(\omega_+)}} \simeq 
		f(\omega_A) 
		+ \frac{\bkappa\cdot \bp_0}{M c} \big( f(\omega_A) + \omega_A f'(\omega_A) \big) 
		- \frac{\hbar \omega_A}{2 Mc^2} \big( 2 f(\omega_A) + \omega_A f'(\omega_A) \big)
\end{equation}
to obtain
\begin{gather}
	\int_0^\infty \dd \omega\; f_\Gamma(\omega) \delta(\widetilde{\omega}_\bk)
		= \omega_A^3 \left( 1 - 3 \tfrac{\hbar \omega_A}{2 M c^2} + 4 \tfrac{(\bp_0 \cdot \bkappa)}{Mc}  \right) \left( d^2 - (\bd\cdot\bkappa)^2\right)
		 - \tfrac{2 \omega_A^3}{Mc} \left( d^2 (\bp_0 \cdot \bkappa) - (\bp_0 \cdot \bd)(\bd\cdot\bkappa)\right) \,,
	\\	 
	\int_0^\infty \dd \omega\; \bkappa\; f_{\dot{\vct{P}}}(\omega) \delta(\widetilde{\omega}_\bk)
		= \omega_A^4 \bkappa \left( 1 - 4 \tfrac{\hbar \omega_A}{2 M c^2} + 5 \tfrac{(\bp_0 \cdot \bkappa)}{Mc}  \right) \left( d^2 - (\bd\cdot\bkappa)^2\right)
		 - \tfrac{2 \omega_A^4}{Mc} \bkappa \left( d^2 (\bp_0 \cdot \bkappa) - (\bp_0 \cdot \bd)(\bd\cdot\bkappa)\right) \,.
\end{gather}
Terms odd in $\bkappa$ vanish in the following integration $\int_{4\pi} d\bkappa$ where we get
\begin{align}
	\int_{4\pi} \dd\bkappa \, \big(d^2 - (\bd \cdot\bkappa)^2 \big) 
		&= \tfrac{8\pi}{3} d^2 \,,
	\\
	5 \int_{4\pi} \dd\bkappa \, \bkappa (\bkappa \cdot \bp_0) \big(d^2 - (\bd \cdot\bkappa)^2 \big)
		&= \tfrac{8\pi}{3} \big( 2 d^2 \bp_0 - (\bp_0\cdot \bd) \bd \big) \,,
	\\
	2 \int_{4\pi} \dd\bkappa \, \bkappa \left( d^2 (\bkappa \cdot \bp_0) - (\bkappa \cdot \bd) (\bd \cdot \bp_0)\right)
		&= \tfrac{8\pi}{3} \big( d^2 \bp_0 - (\bp_0\cdot \bd) \bd \big) \,,
\end{align}
which can be checked, for instance, by choosing $\bp_0 = (0,0,p_0)$ and $\bd=(d_x,d_y,d_z)$.

We thus end up with
\begin{gather}
	\Gamma = \frac{\omega_A^3 d^2}{3 \pi \varepsilon_0 \hbar c^3} \left(1-3\frac{\hbar \omega_A}{2 M c^2}\right) \,,
	\\
	\tder{t}{}\ew{\vct{P}(t)} = - \frac{\omega_A^3 d^2}{3 \pi \varepsilon_0 \hbar c^3} \frac{\hbar \omega_A}{M c^2} \bp_0 \,.
\end{gather}
This shows that the decay rate is independent of the velocity (at least to first order in $v_0/c = \bp_0/(Mc)$) while the momentum changes as given in equ.~(16).
\end{document}